\documentclass{article}
\usepackage{graphicx}

\newcommand{\ket}[1]{\left| #1 \right\rangle}
\newcommand{\bra}[1]{\left\langle #1 \right|}

\newcommand{\Hi}{{\cal H}}

\title{Detecting entanglement with non-hermitian operators}
\author{Mark Hillery$^{(1)}$, Ho Trung Dung$^{(1,2)}$, and Julien Niset$^{(1)} $\\
$^{(1)}$Department of Physics, Hunter College of CUNY \\ 695 Park Avenue \\ New York, NY 10065 \and
$^{(2)}$ Institute of Physics, Academy of Sciences and Technology \\ 1 Mac Dinh Chi Street,
District 1 \\ Ho Chi Minh City, Vietnam}
\begin{document}
\maketitle
\begin{abstract}
We derive several entanglement conditions employing non-hermitian operators.  We start with
two conditions that were derived previously for field mode operators, and use them to derive
conditions that can be used to show the existence of field-atom entanglement and entanglement
between groups of atoms.  The original conditions can be strengthened by making them invariant
under certain sets of local unitary transformations, such as Gaussian operations.  We then apply
these conditions to several examples, such as the Dicke model.  We conclude with a short
discussion of how local uncertainty relations with non-hermitian operators can be used to derive
entanglement conditions.
\end{abstract}

\section{Introduction}
Entanglement has shown itself to be a valuable resource in quantum information in applications
ranging from communication protocols, such as dense coding, to quantum computing.  This has
led to a substantial effort to understand and characterize entanglement (see \cite{horodecki} and
\cite{guehne} for recent reviews).  There is no simple universal test that enables one to tell whether
a given state is entangled, but there are many sufficient conditions.  These include, for example,
entanglement witnesses. An entanglement witness is an operator whose expectation value is nonnegative for separable states, but its expectation value can also be negative, and the states
for which it is are entangled.  For
continuous-variable systems, there are entanglement criteria involving the expectation values
of powers of creation and annihilation operators \cite{simon}-\cite{shchukin}.  Here we would
like to expand upon the work in \cite{hillery} to find stronger entanglement conditions, and to
find conditions for entanglement not just between field modes, but between atoms and field
modes or between groups of atoms.

In \cite{hillery} two conditions that enable one to determine whether the modes in a two-mode state  are entangled were derived.  These conditions are that the modes are entangled if either
\begin{equation}
\label{ineq1}
|\langle A^{\dagger}B\rangle |^{2} > \langle A^{\dagger}AB^{\dagger}B\rangle ,
\end{equation}
or
\begin{equation}
\label{ineq2}
|\langle AB \rangle |^{2} > \langle A^{\dagger}A\rangle \langle B^{\dagger}B\rangle ,
\end{equation}
where $A$ is any power of the annihilation operator for the first mode and $B$ is any power of the
annihilation operators for the second mode.  These conditions are sufficient, but not necessary,
to demonstrate entanglement, that is if either one is satisfied, the state is entangled, but if neither
one is satisfied, we cannot say anything about the entanglement of the state.

The derivations of these inequalities made no use of the special properties of annihilation
operators. In fact, if we have a bipartite system described by a Hilbert space
$\mathcal{H}=\mathcal{H}_{a} \otimes \mathcal{H}_{b}$, and $A$ is any operator on $\mathcal{H}_{a}$
and $B$ is any operator on $\mathcal{H}_{b}$, then if a state satisfies either of the above
conditions, then that state is entangled.  If $A$ or $B$ is hermitian, these conditions are
useless, because the Schwarz inequality guarantees that they will be violated for all states.
Consequently, we will be studying the detection of entanglement with non-hermitian operators.

There have been some previous studies of entanglement making use of nonhermitian operators.
Toth, Simon and Cirac derived an inequality based on uncertainty relations with the number
operator and the mode annihilation operator that can detect entanglement between modes
\cite{toth}. As was mentioned in the previous paragraph,  Zubairy and one of us derived a set of
entanglement conditions based on mode creation and annihilation operators, and a very general
set of conditions based on the same operators was derived by Shchukin and Vogel \cite{shchukin}.

We will also show how our original entanglement conditions can be strengthened by making
use of local unitary invariance.  In the next section this will be studied for some simple, low-dimesnional
systems.  We then go on to look at an atom, or atoms coupled to a single-mode field and develop
entanglement conditions for the field-atoms system that are invariant under Gaussian transformations
on the field mode.  In the following section, we demonstrate that all of the entanglement conditions
we find can be derived from the partial transpose condition, i.e. that if the partial transpose of a
state is not positive, then the original state is entangled.  However, the conditions we derive are
much easier to use than the partial transpose condition itself.  We then go on to study the application
of invariance under Gaussian operations to entanglement conditions for two modes.  We then
examine entanglement in two extended examples, the Dicke model and light modes coupled by
two beam splitters.  In both cases we find a connection between sub-Poissonian statistics of an
input light mode and subsequent entanglement in the system.  Finally, we show how a non-hermitian
version of local uncertainty relations can be used to derive further entanglement conditions between
field modes, and between a collection of atoms and a field mode.

\section{Correlated subspaces}
As mentioned in the Introduction,
another element we will be using in our study of entanglement is its invariance under local unitary
operations.  Because of this invariance, it would be nice if our criteria for entanglement also
possessed this invariance.  As we shall see, we cannot always completely satisfy this condition,
but we can often satisfy it for some subset of local unitary operators.

Let us begin by considering the first inequality above.  Let $|\alpha_{1}\rangle_{a}$ and
$|\alpha_{2}\rangle_{a}$ be two orthogonal vectors in $\mathcal{H}_{a}$ and $|\beta_{1}\rangle_{b}$
and $|\beta_{2}\rangle_{b}$ be two orthogonal vectors in $\mathcal{H}_{b}$.  Now let
$A=|\alpha_{2}\rangle\langle\alpha_{1}|$ and $B=|\beta_{1}\rangle\langle\beta_{2}|$, and consider
the state
\begin{equation}
|\psi\rangle = c_{1}|\alpha_{1}\rangle_{a}|\beta_{1}\rangle_{b} + c_{2} |\alpha_{2}\rangle_{a}
|\beta_{2}\rangle_{b}  ,
\end{equation}
where $|c_{1}|^{2} + |c_{2}|^{2} = 1$.  We find that
\begin{equation}
 |\langle A^{\dagger}B\rangle |^{2} = |c_{1}c_{2}|^{2}, \hspace{5mm}
 \langle A^{\dagger}AB^{\dagger}B\rangle = 0,
\end{equation}
so that Eq.\ ({\ref{ineq1}) is a good way to detect the entanglement of $|\psi\rangle$.

It will still detect it if we add noise to the state.  Defining
\begin{equation}
  P_{\alpha} = \sum_{j=1}^{2}|\alpha_{j}\rangle\langle\alpha_{j}|, \hspace{5mm}
  P_{\beta} =\sum_{j=1}^{2}|\beta_{j}\rangle\langle\beta_{j}| ,
\end{equation}
and considering the density matrix
\begin{equation}
\rho = s|\psi\rangle\langle\psi | +\frac{1-s}{4}P_{\alpha}\otimes P_{\beta} ,
\end{equation}
where $0\leq s \leq 1$, we find that Eq.\ (\ref{ineq1}) shows that the state is entangled if
\begin{equation}
s > \frac{(1+ 16|c_{1}c_{2}|^{2})^{1/2} -1}{8|c_{1}c_{2}|^{2}} .
\end{equation}
If $c_{1}=c_{2}=1/\sqrt{2}$ this gives $s>(\sqrt{5}-1)/2$ and if $|c_{1}c_{2}| \ll 1$, then we have
$s> 1-4|c_{1}c_{2}|^{2}$.  Another possible density matrix that can be detected by this condition
is one of the form
\begin{equation}
\rho = s|\psi\rangle\langle \psi | + \rho_{0}  ,
\end{equation}
where the vectors $|\alpha_{j}\rangle_{a}|\beta_{k}\rangle_{b}$, for $j,k=1,2$ are in the null space
of $\rho_{0}$, and ${\rm Tr}(\rho_{0})=1-s$.  For this density matrix, Eq.\ (\ref{ineq1}) will show that
it is entangled if $s>0$.

Now let us go to a more complicated situation.  Let $S_{1}$ be the linear span of the vectors $\{
|\alpha_{1}\rangle , |\alpha_{2}\rangle \}$, and $S_{2}$ be the linear span of the vectors $\{
|\alpha_{3}\rangle , |\alpha_{4}\rangle \}$, where the vectors $\{ |\alpha_{j}\rangle \in
\mathcal{H}_{a} |j=1, \dots ,4\}$ form an orthonormal set.  We want an entanglement condition that
will detect entanglement in vectors of the form
\begin{equation}
|\psi\rangle = \frac{1}{\sqrt{2}} (|v_{1}\rangle_{a}|\beta_{1}\rangle_{b} + |v_{2}\rangle_{a}
|\beta_{2}\rangle_{b}) ,
\end{equation}
where $|v_{1}\rangle \in S_{1}$ and $|v_{2}\rangle \in S_{2}$.  We would also like the condition to
be independent of the specific vectors $|v_{1}\rangle_{a}$ and $|v_{2}\rangle_{a}$.  One way to
approach this is to make use of the fact that entanglement is invariant under local unitary
transformations. Suppose that $U_{a}$ is a unitary operator on $\mathcal{H}_{a}$ that leaves
$S_{1}$ and $S_{2}$ invariant. For any two vectors in $S_{1}$, $|v_{1}\rangle_{a}$ and
$|v_{1}^{\prime}\rangle_{a}$, and two vectors in $S_{2}$, $|v_{2}\rangle_{a}$ and
$|v_{2}^{\prime}\rangle_{a}$, we can find a $U_{a}$ such that $U_{a}|v_{1}\rangle_{a} =
|v_{1}^{\prime}\rangle_{a}$ and $U_{a}|v_{2}\rangle_{a} = |v_{2}^{\prime}\rangle_{a}$.  Therefore,
if we have an entanglement condition that is invariant under the transformations, $U_{a}$, we will
have one that is independent of the vectors $|v_{1}\rangle_{a}$ and $|v_{2}\rangle_{a}$ in the
equation for $|\psi\rangle$ above.  We can find such a condition by noting that if we have an
operator of the form $|\alpha_{3}\rangle_{a}\langle\alpha_{1}|$ that maps $S_{1}$ to $S_{2}$, then
$U_{a}|\alpha_{3}\rangle_{a}\langle\alpha_{1}|U_{a}^{-1}$ will be a linear combination of the
operators $|\alpha_{j}\rangle_{a}\langle\alpha_{k}|$, where $j=3,4$ and $k=1,2$.  Therefore, let us
choose
\begin{eqnarray}
A  & = & z_{1}|\alpha_{3}\rangle_{a}\langle\alpha_{1}| + z_{2}|\alpha_{4}\rangle_{a}\langle\alpha_{1}|
+ z_{3}|\alpha_{3}\rangle_{a}\langle\alpha_{2}|   \nonumber \\
& & + z_{4}|\alpha_{4}\rangle_{a}\langle\alpha_{2}| ,
\end{eqnarray}
where the complex numbers $z_{1}, \ldots ,z_{4}$ are arbitrary.  The operator B is as before.  We
then have that
\begin{equation}
|\langle A^{\dagger}B\rangle |^{2} - \langle A^{\dagger}AB^{\dagger}B\rangle = \sum_{j,k=1}^{4}
z_{j}^{\ast}M_{jk}z_{k}  ,
\end{equation}
where the $4\times 4$ matrix $M$ depends on the state being considered.  If $M$ has a positive
eigenvalue, then by choosing the $z_{j}$, $j=1,\ldots ,4$ to be the components of the corresponding
eigenvector, we will have an operator $A$ that shows that the state we are considering is
entangled.

Let us carry this out explicitly for the density matrix
\begin{equation}
\rho = s|\psi\rangle\langle\psi |+ \frac{1-s}{8} P_{\alpha}\otimes P_{\beta} ,
\end{equation}
where $P_{\alpha}$ is the projection onto $S_{1}\cup S_{2}$ and $P_{\beta}$ is as before.  We then
find that
\begin{equation}
M_{jk} =\frac{s^{2}}{4} \eta_{j}\eta_{k}^{\ast} - \frac{1-s}{8} \delta_{jk} ,
\end{equation}
where
\begin{eqnarray}
 \eta_{1} = \langle v_{1}|\alpha_{1}\rangle \langle\alpha_{3}|v_{2}\rangle, &
 \eta_{2} = \langle v_{1}|\alpha_{1}\rangle \langle\alpha_{4}|v_{2}\rangle,  \nonumber \\
 \eta_{3} = \langle v_{1}|\alpha_{2}\rangle \langle\alpha_{3}|v_{2}\rangle, &
 \eta_{4} = \langle v_{1}|\alpha_{2}\rangle \langle\alpha_{4}|v_{2}\rangle .
\end{eqnarray}
We can then express $M$ as
\begin{equation}
M = \frac{s^{2}}{4} |\eta\rangle\langle \eta |- \frac{1-s}{8} ,
\end{equation}
where the vector $|\eta\rangle$ has components given by $\eta_{j}$, $j=1,\ldots ,4$.  It is then
clear that $M$ has three negative eigenvalues, corresponding to directions orthogonal to
$|\eta\rangle$, with the remaining eigenvalue equal to $(2s^{2} + s  -1)/8$.  This is positive if
$s > 1/2$, and therefore, the state is entangled if $s > 1/2$.    Note that this condition is
independent of $|v_{1}\rangle_{a}$ and $|v_{2}\rangle_{a}$.

\section{Two-level atom coupled to the field}
The above entanglement condition can be applied to a two-level atom coupled to a single-mode field.
The atom can either absorb a photon and go from its lower to its upper state, or emit a photon and
go from its upper to its lower state. Let the lower state of the atom be $|g\rangle$ and the
excited state be $|e\rangle$, and let us consider photon states with $0$, $1$, $2$, and $3$
photons.  If we start the atom in its excited state and a superposition of $0$, and $2$ photons, as
the system evolves there will be some amplitude to be in these states, but there will also be
amplitudes to be in the states $|g\rangle |1\rangle$ and $|g\rangle |3\rangle$.  The upper state of
the atom will be correlated with photon subspace spanned by $|1\rangle$ and $|3\rangle$, and the
lower state will be correlated with the subspace spanned by $|0\rangle$ and $|2\rangle$.
Therefore, the entanglement conditions developed in the previous section could be used to test for
entanglement in this system.

In considering an atom interacting with a single-mode field, we do not usually confine our attention
to states of three photons or fewer, so a different set of entanglement conditions could be more
useful.  Instead of limiting the photon number, we will consider all possible photon states.  In that
case we will have to give up invariance of the conditions under all unitary transformations of the
photon Hilbert space.  We can, however, derive some simple conditions if we restrict the set of
unitary transformations to those corresponding to Gaussian operations.  These operations consist
of translations, rotations and squeezing transformations.  In particular, if $a$ and $a^{\dagger}$ are
the annihilation and creation operators for the mode, the translations are given by the operators
$D(\alpha )=\exp (\alpha a^{\dagger} - \alpha^{\ast} a)$, where $\alpha$ is an arbitrary complex
number, rotations are given by $R(\theta )= \exp (i\theta a^{\dagger}a)$, where $0\leq \theta <2\pi$,
and the squeezing transformation is given by $S(z) = \exp [ ( z^{\ast}a^{2} -z(a^{\dagger})^{2})/2 ] $,
where $z=r\exp (i\phi )$ is a complex number.  These transformations act as follows:
\begin{eqnarray}
D(\alpha )^{\dagger} a D(\alpha ) & = & a + \alpha, \nonumber \\
R(\theta )^{\dagger} a R(\theta ) & = & e^{i\theta }a,  \nonumber \\
S(z)^{\dagger} a S(z) & = & a \cosh r - a^{\dagger} e^{-i\phi} \sinh r  .
\end{eqnarray}
Note that these transformations send the creation and annihilation operators into a linear combinations
of the annihilation operator, the creation operator, and a constant.

In order to find an entanglement condition that is invariant under Gaussian transformations of the
field mode, we set
\begin{eqnarray}
 A & = & \sigma^{-} = |g\rangle \langle e|, \nonumber \\
 B & = & z_{1}(a-\langle a\rangle ) + z_{2} (a^{\dagger} - \langle a^{\dagger}\rangle ) ,
\end{eqnarray}
and substitute these into Eq.\ (\ref{ineq1}).  The result is
\begin{equation}
\label{at-field}
\sum_{j,k=1}^{2}z_{j}^{\ast}M_{jk}z_{k} > 0 ,
\end{equation}
where now
\begin{equation}
 M = \left( \begin{array}{cc} |\langle \sigma^{+} \Delta a\rangle |^{2}
 -\langle P_{e}(\Delta a)^{\dagger}\Delta a\rangle &
 \langle \sigma^{-}(\Delta a)^{\dagger}\rangle\langle \sigma^{+}(\Delta a)^{\dagger}\rangle
 - \langle P_{e}( (\Delta a)^{\dagger})^{2}\rangle \\
 \langle\sigma^{-}\Delta a\rangle \langle\sigma^{+}\Delta a\rangle
 - \langle P_{e}(\Delta a)^{2}\rangle  &
 |\langle \sigma^{-}\Delta a\rangle |^{2} - \langle P_{e}\Delta a(\Delta a)^{\dagger}\rangle
\end{array} \right) ,
\end{equation}
where $\Delta a = a - \langle a\rangle$, $\sigma^{+}=|e\rangle\langle g|$ and $P_{e}=
|e\rangle\langle e|$.  If we transform the state of the field mode with a Gaussian transformation,
the effect in Eq.\ (\ref{at-field}) is just to change the values of $z_{1}$ and $z_{2}$.

Now if we can find a value of $z_{1}$ and $z_{2}$ so that Eq.\ (\ref{at-field}) is satisfied, then the
state is entangled.  This will be possible if the matrix $M$ has a positive eigenvalue.  Therefore, our
entanglement condition, which is now invariant under Gaussian transformations of the field mode, is
that the state is entangled if $M$ has a positive eigenvalue.  That means, for example, that it can
detect entanglement in states of the form $I_{at}\otimes S(z)D(\alpha)(|e\rangle |0\rangle + |g\rangle
|1\rangle )/\sqrt{2}$, where $I_{at}$ is the identity operator for the atomic system, for any value of
$\alpha$ and $z$.

Let us now look at two examples of atom-field entanglement.
Let us first consider a two-level atom in the rotating wave approximation interacting with a single field mode,
which is initially in a thermal state.  On resonance,
the Hamiltonian for this system is
\begin{equation}
 H = \omega a^{\dagger}a + \frac{\omega}{2} \sigma^{z}
 +  \kappa (\sigma^{+}a + \sigma^{-}a^{\dagger}) .
\end{equation}
The thermal state density matrix is given by
\begin{equation}
 \rho_{therm} = \frac{1}{(1+\bar{n})} \sum_{n=0}^{\infty} \left( \frac{\bar{n}}{1+\bar{n}}\right)^{n}
 |n\rangle\langle n| .
\end{equation}
With this initial state we find that the off-diagonal elements of $M$ are initially zero, and remain zero
for all times.    The element in the lower right-hand corner ($M_{22}$) is always negative, so only
the element in the upper left-hand corner ($M_{11}$) can possibly become positive.  We plot
$M_{11}$ for several different values of $\bar{n}$ in Figure 1.  As can be seen, for sufficiently small
values of $\bar{n}$, our condition shows that there are times at which the atom and the field mode
are entangled.

\begin{figure}
\includegraphics{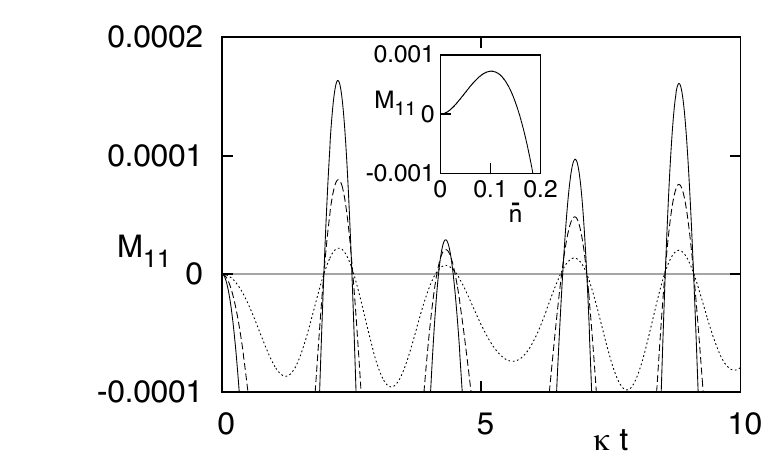}
\caption{The eigenvalue $M_{11}$ as a function of time for average
 photon numbers $\bar{n}=0.03$ (solid line), $\bar{n}=0.02$ (dashed line),
 and $\bar{n}=0.01$ (dotted line).
 The inset shows $M_{11}$ as a function of $\bar{n}$ at $\kappa t=2.24$.}
 \end{figure}

 Let us now look at two two-level atoms interacting on resonance with a single mode field.  The
 Hamiltonian of the system is now
\begin{equation}
 H=\omega a^\dagger a + \frac{\omega}{2}(\sigma^z_1 + \sigma^z_2) + \kappa \big( ( a \sigma^+_1
 + a^\dagger \sigma^-_1)+ ( a \sigma^+_2 + a^\dagger \sigma^-_2) \big) .
\end{equation}
This Hamiltonian preserves the total number of excitations of the system, so we can express the
total Hilbert space of the system as a direct sum of subspaces containing fixed numbers of
excitations, i.e. $ \Hi=\oplus\Hi^{(n)}$ .  For $n>1$ the subspace $\Hi^{(n)}$ is spanned by the
vectors $|n,g,g\rangle$, $|n-1,g,e\rangle$, $|n-1,e,g\rangle$, and $|n-2,e,e\rangle$, where the
first slot contains the photon number, and the remaining two slots contain the states of the atoms.
For $n=1$ it is spanned by $|1,g,g\rangle$, $|0,g,e\rangle$, and $|0,e,g\rangle$.  If we start in
the state $|n,g,g\rangle$ at time $t=0$, we find that $M_{12}=M_{21}=0$ for all times, and
$M_{22}\leq 0$.  Thus our entanglement condition reduces to $M_{11}>0$.  Finding an explicit
expression for $M_{11}$ (see the Appendix) this condition becomes
\begin{eqnarray}
\label{2at0}
  &&\frac{n}{2n-1} \sin^2(\Omega t)[\cos (\Omega t) + 2(n-1)]^2
\nonumber\\
  && \hspace{1cm}- (n-1)
  \Bigl[ 2(n-2)(\cos (\Omega t) -1)^2 + (2n-1)\sin^2(\Omega t) \Bigr] >0,
\end{eqnarray}
where $\Omega = \kappa\sqrt{2(2n-1)}$.
It is worthwhile to look at some special cases. 
For $n=1$ and $n=2$,  the condition reads as
\begin{eqnarray}
\label{n_1}
  &&\sin^2(\Omega t)\cos^2(\Omega t)>0,
\\
  && |\cos(\Omega t) + 2 | > \frac{3}{\sqrt{2}}\ ,
\end{eqnarray}
respectively. For the single-photon case, it is not known which of the two atoms absorbs the
photon, so that the atoms are always entangled. The condition becomes more restrictive as $n$
increases.
For times in the neighborhood of
 $2k\pi/\Omega$
\begin{equation}
\label{time}
 t = \epsilon + k \frac{2 \pi}{\Omega},
\end{equation}
 where $\epsilon$ is small but nonvanishing and $k$ is an integer, one
 can make an $\epsilon$-expansion of the left-hand side of Eq. (\ref{2at0}) around $0$
 to obtain
\begin{equation}
\label{large_n_a}
       (2n-1)\epsilon^2 - \frac{1}{6}(3n^2+n+4) \epsilon^4 + O(\epsilon^6) >0.
\end{equation}
This can be satisfied for an arbitrarily large $n$, provided $\epsilon$ is small enough, that is
there always exists a time interval about $2k\pi/\Omega$ during which entanglement is detected.
Obviously, for an increasing $n$, this time interval is reduced.

We can also study the entanglement between both atoms and the field.   Now we choose $A=a$ and
$B=J^-$ with $J^- = \sigma_1^- + \sigma_2^-$ in Eq.\ (\ref{ineq1}). The condition for entanglement
becomes
\begin{equation}
 |\langle a^\dagger J^-\rangle|^2 - \langle a^\dagger a J^+ J^-\rangle > 0,
\end{equation}
or, by using the state at time $t$ as given in the Appendix,
\begin{eqnarray}
\label{2at}
  &&\frac{n}{2n-1} \sin^2(\Omega t)[\cos (\Omega t) + 2(n-1)]^2
\nonumber\\
  && \hspace{1cm}- (n-1)
  \Bigl[ (n-2)(\cos (\Omega t) -1)^2 + (2n-1)\sin^2(\Omega t) \Bigr] >0.
\end{eqnarray}
For $n=1$ and $n=2$ the above equation is the same as Eq. (\ref{2at0}). In general, this condition
is easier to satisfy than the condition (\ref{2at0}) due to the fact that its second term is smaller than
its counterpart in Eq. (\ref{2at0}). As a consequence, there are moments when we detect
 entanglement between the field and both atoms, while no entanglement is detected between
 the field and a single atom. For times about $2k\pi/\Omega$, see Eq. (\ref{time}), one can expand
 (\ref{2at}) with respect to $\epsilon$ around $0$
\begin{equation}
\label{large_n_f}
       (2n-1)\epsilon^2 - \frac{1}{12}(3n^2+11n+2) \epsilon^4 + O(\epsilon^6) >0.
\end{equation}
 Again for large $n$ one can detect
 entanglement during time windows centered around $2k\pi/\Omega$.

\section{Relation to partial transpose condition}
The derivations of the entanglement conditions, Eqs.\ (\ref{ineq1}) and (\ref{ineq2}),
which were presented in \cite{hillery}, did not make use of the Positive Partial Transpose (PPT)
condition.  This condition states that the partial transpose of a separable density matrix is a
positive operator, which implies that if the partial transpose of a density matrix is not positive,
the original density matrix was entangled.
However, it was subsequently shown that in the case that $A$ and $B$ are powers of
mode creation and annihilation operators, the conditions are a consequence of the PPT condition
\cite{nha}.  We will now show that this is true in general, i.e. that no matter what the choice of
$A$ and $B$, the entanglement conditions in Eqs.\ (\ref{ineq1}) and (\ref{ineq2}) are consequences
of the PPT condition.

Let us begin by discussing one of these conditions.  Our system is divided into subsystems $a$
and $b$, and let $A$ be an operator acting on subsystem $a$ and $B$ an operator acting on
subsystem $b$.  A separable state must satisfy the condition \cite{hillery}
\begin{equation}
\label{sepcond1}
|\langle A^{\dagger}B\rangle |^{2} \leq \langle A^{\dagger}AB^{\dagger}B\rangle .
\end{equation}
If this condition is violated, the state is entangled.
Now let us see if we can derive this condition from the PPT condition.  We start with the condition
\begin{equation}
\label{schwarzcond}
|\langle AB\rangle |^{2} \leq  \langle A^{\dagger}AB^{\dagger}B\rangle ,
\end{equation}
which follows from the Schwarz inequality.  These expectation values are taken with respect to a
density matrix $\rho$, and the matrix element of this density matrix are
\begin{equation}
\rho_{m,\mu ; n,\nu} = (\,_{a}\langle m|\,_{b}\langle \mu |) \rho ( |n\rangle_{a} |\nu\rangle_{b}) ,
\end{equation}
where $\{ |m\rangle_{a} \}$ is an orthonormal basis for $a$ and $\{ |\mu \rangle_{b} \}$ is an orthonormal
basis for $b$.  Now define the partial transpose of $\rho$ with respect to subsystem $a$, $\rho^{T_a}$,
to be the operator with matrix elements
\begin{equation}
\rho^{T_a}_{m,\mu ; n,\nu}  = \rho_{n,\mu ; m,\nu} .
\end{equation}
For a separable density matrix, $\rho^{T_a}$ should also be a valid density matrix, that is it should
be positive and have a trace equal to one.  That implies that the inequality in Eq.\ (\ref{schwarzcond})
will hold if the expectation values are taken with respect to $\rho^{T_a}$.  We have that
\begin{eqnarray}
\label{partrans}
{\rm Tr}(AB\rho^{T_a}) & = & \sum_{m,n} \sum_{\mu ,\nu} A_{nm}B_{\nu \mu} \rho^{T_a}_{m\mu ; n\nu}
\nonumber \\
 & = & \sum_{m,n} \sum_{\mu ,\nu} A_{nm}B_{\nu \mu} \rho_{n,\mu ; m,\nu}   \nonumber  \\
  & = &\sum_{m,n} \sum_{\mu ,\nu} (A_{mn}^{\dagger})^{\ast} B_{\nu \mu}  \rho_{n,\mu ; m,\nu} .
 \end{eqnarray}
 If $A_{mn}^{\dagger}$ is real, then we see that ${\rm Tr}(AB\rho^{T_a})={\rm Tr}(A^{\dagger}B\rho)$.
 This would be the case, for example, if $A$ were a product of mode creation and annihilation
 operators and the basis is the number-state basis.  Similarly, we find that if $(A^{\dagger}A)_{mn}$
 is real, then ${\rm Tr}(A^{\dagger}AB^{\dagger}B\rho^{T_a})={\rm Tr}(A^{\dagger}AB^{\dagger}B\rho)$.
 If the density matrix is separable and all of the reality conditions are satisfied, then we have that
 Eq.\ (\ref{schwarzcond}), with the expectation values taken with respect to $\rho^{T_a}$, implies
 Eq.\ (\ref{sepcond1}), with the expectation values taken with respect to $\rho$.  Under these
 conditions, the inequality in Eq.\ (\ref{sepcond1}) follows from the PPT condition.

 As we can see, the above derivation depends on the fact that certain matrix elements are real.  The
 condition can be relaxed a bit; as long as the matrix elements $A_{mn}^{\dagger}$ all have the
 same phase, the derivation will work.  However, the original derivation of the Eq.\ (\ref{sepcond1})
 did not require these conditions on the matrix elements.

 It is the none the less the case that Eq.\ (\ref{sepcond1}) can be derived from the PPT condition
 with no additional assumptions.
 Consider two operators $F$ acting on $\mathcal{H}_{a}$ and $B$ acting on $\mathcal{H}_{b}$.
 We have that, as in the previous paragraph, for a separable state
 \begin{equation}
 |\langle FB\rangle |^{2} \leq  \langle F^{\dagger}FB^{\dagger}B\rangle .
\end{equation}
Now, from above, we see that
\begin{eqnarray}
{\rm Tr}(FB \rho^{T_a} ) & = &\sum_{m,n} \sum_{\mu ,\nu} (F_{mn}^{\dagger})^{\ast} B_{\nu \mu}
\rho_{n,\mu ; m,\nu},   \\
{\rm Tr}( F^{\dagger}FB^{\dagger}B  \rho^{T_a} ) & = & \sum_{m,n} \sum_{\mu ,\nu}
[(F^{\dagger}F)_{mn}]^{\ast} (B^{\dagger}B)_{\nu \mu}  \rho_{n,\mu ; m,\nu}.
\end{eqnarray}
Define the operator $G$, acting on $\mathcal{H}_{a}$, by
\begin{equation}
G=\sum_{m,n} F^{\ast}_{mn} |m\rangle \langle n| .
\end{equation}
The quantities involving the partially transposed density matrix can then be expressed as
\begin{eqnarray}
{\rm Tr}(FB \rho^{T_a} ) & = & \langle G^{\dagger}B\rangle, \\
{\rm Tr}( F^{\dagger}FB^{\dagger}B  \rho^{T_a} ) & = & \langle  G^{\dagger}GB^{\dagger}B \rangle .
\end{eqnarray}
For any separable density matrix, we must have
\begin{equation}
| \langle G^{\dagger}B\rangle |^{2} \leq \langle  G^{\dagger}GB^{\dagger}B \rangle .
\end{equation}
In order to recover our original inequality, choose
\begin{equation}
F=\sum_{m,n} (A_{mn})^{\ast} |m\rangle \langle n| ,
\end{equation}
which implies that $G=A$, and this completes the derivation of the inequality from the PPT condition.
The derivation, for separable states, of the condition
\begin{equation}
|\langle AB \rangle |^{2} \leq \langle A^{\dagger}A\rangle \langle B^{\dagger}B\rangle ,
\end{equation}
using the PPT condition, is similar.

\section{Two modes}
Let us now look at the case of two modes with annihilation operators $a$ and $b$.  We will first
find an entanglement condition that is invariant under Gaussian transformations of one of the
modes.  Let us set
 $A=z_{1}(\Delta a)^{\dagger} + z_{2}\Delta a $
 and $B=b$.  Then the condition
 $|\langle A^{\dagger}B\rangle|^{2} > \langle A^{\dagger}AB^{\dagger}B\rangle$ can be written as
\begin{equation}
\langle v|M|v\rangle >0 ,
\end{equation}
where $|v\rangle$ is the two-component vector
\begin{equation}
|v\rangle = \left( \begin{array}{c} z_{1} \\  z_{2} \end{array} \right) ,
\end{equation}
and $M$ is the $2\times 2$ matrix
\begin{equation}
 M=\left( \begin{array}{cc} |\langle \Delta a b\rangle |^{2}
 - \langle \Delta a(\Delta a)^{\dagger}b^{\dagger}b\rangle &
 \langle \Delta ab\rangle \langle (\Delta a)^{\dagger}b\rangle^{\ast}
 - \langle (\Delta a)^{2}b^{\dagger}b\rangle \\
 \langle (\Delta a)^{\dagger}b\rangle \langle \Delta ab\rangle^{\ast}
 - \langle (\Delta a^{\dagger})^{2}b^{\dagger}b\rangle &
 |\langle (\Delta a^{\dagger}b\rangle |^{2}
 - \langle (\Delta a)^{\dagger}\Delta ab^{\dagger}b\rangle \end{array} \right) .
\end{equation}
In this form, we can see that if $M$ has at least one positive eigenvalue, we can find a vector
$|v\rangle$ so that the entanglement condition $\langle v|M|v\rangle >0$ is satisfied, so
that we can say that a state is entangled if the matrix $M$ that results from it has a positive
eigenvalue.  As before, this condition  is invariant under Gaussian operations,
and, consequently, so is our entanglement condition.  For this simple $2\times 2$ case, we can make
the condition more explicit.  Denoting the matrix elements of $M$ by $M_{jk}$, where $j,k=1,2$, the
larger of the two eigenvalues is
\begin{equation}
\lambda = \frac{1}{2}\{ (M_{11}+M_{22}) + [(M_{11}-M_{22})^{2}+ 4|M_{12}|^{2}]^{1/2} \} ,
\end{equation}
and the condition that $\lambda >0$ is either that $(M_{11}+M_{22}) > 0$, or that
$|M_{12}|^{2} > M_{11}M_{22}$.

As a short example of how our new condition is stronger than the one originally proved in
\cite{hillery}, consider the state
\begin{equation}
|\psi^{\prime}\rangle = \frac{1}{\sqrt{2}}S_{a}(z)\otimes I_{b}(|0\rangle_{a}|1\rangle_{b} + |1\rangle_{a}
|0\rangle_{b}) .
\end{equation}
This state is obtained by applying a Gaussian operation to the state $|\psi\rangle = (|0\rangle_{a}
|1\rangle_{b} + |1\rangle_{a}|0\rangle_{b})/\sqrt{2}$ .  For the state $|\psi\rangle$, the matrix in the
previous paragraph is diagonal, and the lower right-hand element is positive, so that the state
is entangled.  Since $|\psi^{\prime}\rangle$ is obtained from $|\psi\rangle$ by a Gaussian operation,
the criterion in the previous paragraph will also show that $|\psi^{\prime}\rangle$ is entangled
for any $z$.  However, if we apply the criterion
\begin{equation}
|\langle a^{\dagger}b\rangle |^{2} > \langle a^{\dagger}ab^{\dagger}b\rangle ,
\end{equation}
presented in \cite{hillery}, we find that it only shows that the state is entangled if $\tanh |z| < 1/\sqrt{2}$.
Therefore, the new condition is an improvement on the old one.

Now suppose we set
\begin{eqnarray}
A & = & z_{1}(a-\langle a\rangle ) + z_{2} (a^{\dagger} - \langle a^{\dagger}\rangle ), \nonumber  \\
B & = &  w_{1}(b-\langle b\rangle ) + w_{2} (b^{\dagger} - \langle b^{\dagger}\rangle ) .
\end{eqnarray}
Let $\mathcal{H}_{1}$ and $\mathcal{H}_{2}$ be two two-dimensional Hilbert spaces with
\begin{equation}
|u\rangle = \left( \begin{array}{c} z_{1} \\ z_{2} \end{array} \right)  \in \mathcal{H}_{1},
\end{equation}
and
\begin{equation}
|v\rangle = \left( \begin{array}{c} w_{1} \\ w_{2} \end{array} \right) \in \mathcal{H}_{2}  .
\end{equation}
The condition $|\langle A^{\dagger}B\rangle |^{2} > \langle A^{\dagger}AB^{\dagger}B\rangle$ can be
written as $(\langle u|\otimes \langle v|)X(|u\rangle \otimes |v\rangle )> 0$, where $X$ is a linear
hermitian operator on $\mathcal{H}_{1}\otimes \mathcal{H}_{2}$, and this operator depends on
the state being considered.  Therefore, a state is entangled if there exists a product state,
$|u\rangle \otimes |v\rangle$ in $\mathcal{H}_{1}\otimes \mathcal{H}_{2}$, such that
\begin{equation}
\label{prodcond}
(\langle u|\otimes \langle v|)X(|u\rangle \otimes |v\rangle )> 0 .
\end{equation}

One way of determining whether the above condition can be satisfied is by examining the operator
$X_{v}=\langle v|X|v\rangle$, where $|v\rangle$ is a vector in $\mathcal{H}_{2}$ and $X_{v}$ is an
operator in $\mathcal{H}_{1}$.  If for any vector $|v\rangle \in \mathcal{H}_{2}$ the operator
$X_{v}$ has a positive eigenvalue, then the condition in Eq.\ (\ref{prodcond}) can be satisfied.
This can be somewhat tedious since we have to consider all possible vectors $|v\rangle$, so having
some simpler criteria is also useful.

The first criterion  involves the eigenvalues of the reduced matrix
$X_{1}= {\rm Tr}_{2}(X)$.  We will show that if $X_{1}$ has a positive eigenvalue, then a product
vector satisfying the above equation exists.  Suppose that $X_{1}$ has a positive eigenvalue,
$\lambda$, and that the
corresponding eigenstate is $|u\rangle$.  In addition, let $\{ |v_{j}\rangle | j=1,2 \}$ be an orthonormal
basis for $\mathcal{H}_{2}$.  We then have that
\begin{equation}
\langle u|X_{1}|u\rangle = \sum_{j=1}^{2} (\langle u|\otimes \langle v_{j}|) X
(|u\rangle \otimes |v_{j}\rangle ) = \lambda > 0.
\end{equation}
Since the sum in the above equation is positive, at least one of the terms must be positive, i.e.
$ (\langle u|\otimes \langle v_{j}|) X (|u\rangle \otimes |v_{j}\rangle ) >0$ for some $j$.  Therefore,
if $X_{1}$ has a positive eigenvalue, the state is entangled.  This is, however, only a sufficient
condition for a product vector satisfying Eq.\ (\ref{prodcond}) to exist, not a necessary one.  That is,
it is possible for $X_{1}$ to have only negative or zero eigenvalues and still be able to find a product
vector satisfying Eq.\ (\ref{prodcond}).  An example is provided by the operator
\begin{equation}
X=(|01\rangle + |10\rangle )(\langle 01| + \langle 10|) -2(|01\rangle - |10\rangle )
(\langle 01| - \langle 10|) ,
\end{equation}
for which $X_{1}=-I$, but $(\langle +x|\langle +x|)X(|+x\rangle |+x\rangle )>0$.

A second criterion involves the eigenvalues of $X$ itself.  If $X$ has two or more positive
eigenvalues, then we can find a product vector satisfying Eq.\ (\ref{prodcond}).  This can
be seen by showing that a product vector can be found in the subspace spanned by
the eigenvectors corresponding to the two positive eigenvalues.  Let $|x_{1}\rangle$ and
$|x_{2}\rangle$ be the two eigenvectors with positive eigenvalues, and let the
Schmidt decomposition of $|x_{1}\rangle$ be given by
\begin{equation}
|x_{1}\rangle = \sum_{j=1}^{2} \kappa_{j} |\xi_{j}\rangle |\zeta_{j}\rangle .
\end{equation}
We can expand $|x_{2}\rangle$ in the Schmidt basis of $|x_{1}\rangle$ as
\begin{equation}
|x_{2}\rangle = \sum_{j=1}^{2} \sum_{k=1}^{2} d_{jk}  |\xi_{j}\rangle |\zeta_{k}\rangle .
\end{equation}
The vector $\alpha |x_{1}\rangle + \beta |x_{2}\rangle$ is then given by
\begin{equation}
\alpha |x_{1}\rangle + \beta |x_{2}\rangle = \sum_{j=1}^{2} \sum_{k=1}^{2} c_{jk}
|\xi_{j}\rangle |\zeta_{k}\rangle  ,
\end{equation}
where
\begin{equation}
c_{jk} = \alpha \kappa_{1} \delta_{j1}\delta_{k1} + \alpha \kappa_{2} \delta_{j2} \delta_{k2}
+ \beta d_{jk}  .
\end{equation}
Now, $\alpha |x_{1}\rangle + \beta |x_{2}\rangle$ will be a product state if $(c_{11}/c_{12})
=(c_{21}/c_{22})$ ,
and making use of the above equation, this condition can be expressed as
\begin{equation}
\frac{y\kappa_{1} + d_{11}}{d_{12}} = \frac{d_{21}}{y\kappa_{2} + d_{22}} ,
\end{equation}
where $y=\alpha /\beta$.  This leads to a quadratic equation for $y$, which can be solved.
Therefore, there is a product vector in the span of $|x_{1}\rangle$ and $|x_{2}\rangle$.

A simple example of this procedure is given by examining the entanglement of the state
\begin{equation}
\rho = s |\psi_{01}\rangle\langle\psi_{01}| + \frac{1-s}{4} P_{01}^{(a)}\otimes P_{01}^{(b)} ,
\end{equation}
where $|\psi_{01}\rangle = ( |0\rangle_{a}|1\rangle_{b} + |1\rangle_{a} |0\rangle_{b})/\sqrt{2}$,
and $P_{01}^{(a)}$ and $P_{01}^{(b)}$ project onto the zero and one-photon states of the
$a$ and $b$ modes, respectively, i.e. $P_{01}^{(a)} = |0\rangle_{a}\langle 0|
+ |1\rangle_{a}\langle 1|$, and similarly for $P_{01}^{(b)}$.  In Ref. \cite{hillery} it was found that
using the condition in Eq.\ (\ref{ineq1}) with the choice $A=a$ and $B=b$ shows that this state
is entangled for $s>(\sqrt{5}-1)/2$.  With the choice of $A$ and $B$ given above, we find that
the eigenvalues of $X(v)$ are given by the solutions of the equation
\begin{equation}
\lambda^{2} -\left(\frac{s^{2}-2}{4}\right) \lambda + \frac{1}{16}[1-s^{3}(|w_{1}|^{2}-|w_{2}|^{2})^{2}
-2s^{2}(|w_{1}|^{4}+|w_{2}|^{4})] =0 ,
\end{equation}
where we have set $|w_{1}|^{2} + |w_{2}|^{2}=1$, because only the direction of $|v\rangle$ is
important.  One of the eigenvalues will be positive if the last term in this equation is negative, and this
gives us the condition
\begin{equation}
1< s^{3}(|w_{1}|^{2}-|w_{2}|^{2})^{2} +2s^{2}(|w_{1}|^{4}+|w_{2}|^{4}) .
\end{equation}
From this we find that the state is entangled if $s > 0.474$, which is an improvement over
$(\sqrt{5}-1)/2 = 0.618$.

\section{Extended examples}
\subsection{The Dicke model}
We would now like to use the simplest versions of our entanglement conditions to study entanglement
between different subsystems in the Dicke model.  The entanglement of the ground state in the Dicke
model was studied in \cite{buzek}, but we will concentrate on entanglement generated by
the dynamics.  This model consists of $N$ two-level atoms, which
are contained in a volume whose dimensions are small compared to an optical wavelength, coupled
to a single mode of the radiation field.  Assuming again that the interaction is on resonance, and that there is no atom-atom interaction, the Hamiltonian of the system can be written as
\begin{eqnarray}
 H&=&\omega a^\dagger a + \frac{\omega}{2}\sum_{i=1}^N \sigma^z_i
 + \kappa \sum_{i=1}^N ( a \sigma^+_i + a^\dagger \sigma^-_i)
\nonumber \\
&=& \omega a^\dagger a + \omega S^z + \kappa ( a S^+ + a^\dagger S^- ) ,
\end{eqnarray}
after the introduction of the collective spin operators
\begin{eqnarray}
  &&S^+=\sum_{i=1}^N \sigma^+_i ,\qquad S^-=\sum_{i=1}^N \sigma^-_i,
\nonumber \\
  &&S^z=\frac{1}{2}[S^+,S^-].
\end{eqnarray}
The equations of motion resulting from this Hamiltonian are difficult to solve,
so we will introduce an approximation
based on the Holstein-Primakoff representation of the spin operators.
The Holstein-Primakoff transformation expresses the collective spin operators in terms of the bosonic operators $\xi^\dagger$ and $\xi$. If we choose the ground state of these
new operators to correspond to
the atomic state in which all of the atoms are in their lower energy level, the transformation is given by
\begin{eqnarray}
 S^+=\xi^\dagger \sqrt{2S-\xi^\dagger \xi}, \qquad
 S^-=\sqrt{2S-\xi^\dagger \xi}\xi
\end{eqnarray}
with $S=N/2$. From these definitions, it follows that
\begin{eqnarray}
 S^z =-S + \xi^\dagger \xi .
\end{eqnarray}
If the number of excitations of the atomic system remains small with respect to the total number of atoms, i.e. $\langle \xi^\dagger \xi \rangle \ll N$, then we can expand the square roots keeping only the
lowest order terms
\begin{eqnarray}
 S^+\approx \sqrt{N} \xi^\dagger, \qquad S^-\approx \sqrt{N} \xi,
\end{eqnarray}
and the Hamiltonian of the system simplifies to
\begin{equation}
 H=\omega a^\dagger a + \omega \xi^\dagger \xi + \kappa \sqrt{N} (a \xi^\dagger + a^\dagger \xi).
\end{equation}

So far, we have grouped the atoms into a single subsystem.  However, we would like to study the
entanglement between groups of atoms, so we will split them into two groups, each group having
its own collective spin operators.   Let us divide the atoms in two groups, one consisting of $k$
atoms and the other consisting of $N-k$ atoms,
\begin{eqnarray}
 S_1^+&=\sum_{i=1}^k \sigma^+_i\approx\sqrt{k}\xi_1^\dagger, \hspace{10mm}
 S_2^+=\sum_{i=k+1}^N \sigma^+_i\approx\sqrt{N-k}\xi_2^\dagger,
\nonumber\\
 S_1^-&=\sum_{i=1}^k \sigma^-_i\approx\sqrt{k}\xi_1, \hspace{10mm}
 S_2^-=\sum_{i=k+1}^N \sigma^-_i\approx\sqrt{N-k}\xi_2.
\end{eqnarray}
In the Holstein-Primakoff representation with only the lowest order terms retained, the Hamiltonian of
this tripartite system is
\begin{equation}
\label{Ham}
 H=\omega (a^\dagger a + \xi_1^\dagger \xi_1 + \xi_2^\dagger \xi_2)
 + \kappa \sqrt{k} (a \xi_1^\dagger + a^\dagger \xi_1)
 + \kappa \sqrt{N-k} (a \xi_2^\dagger + a^\dagger \xi_2).
\end{equation}

We would now like to diagonalize this Hamiltonian.  In order to do so, we first express it as
\begin{equation}
 H=v^\dagger M v
\end{equation}
with $v=(a,\xi_1,\xi_2)^T$ and
\begin{equation}
 M=\left(
\begin{array}{ccc}
 \omega & \kappa \sqrt{k} & \kappa \sqrt{N-k}\\
\kappa \sqrt{k} & \omega & 0\\
\kappa \sqrt{N-k} & 0 & \omega
\end{array}
\right) .
\end{equation}
The matrix $M$ can be diagonalized by the introduction of the new modes
\begin{eqnarray}
\label{newmodes}
\nonumber b_0 &=& \frac{1}{\sqrt{2}} a - \sqrt{\frac{k}{2N}}\xi_1 - \sqrt{\frac{N-k}{2N}}\xi_2,\\
\nonumber b_1 &=& \sqrt{\frac{N-k}{N}}\xi_1 - \sqrt{\frac{k}{N}}\xi_2,\\
 b_2&=&\frac{1}{\sqrt{2}} a + \sqrt{\frac{k}{2N}}\xi_1 + \sqrt{\frac{N-k}{2N}}\xi_2 ,
\end{eqnarray}
 to which correspond the eigenvalues
\begin{eqnarray}
 &&\lambda_0=\omega - \kappa \sqrt{N}
\nonumber\\
 &&\lambda_1=\omega,
\nonumber \\
 &&\lambda_2=\omega + \kappa \sqrt{N} ,
\end{eqnarray}
respectively. Setting $\Omega = \kappa\sqrt{N}$, the Hamiltonian can now be expressed as
\begin{eqnarray}
\label{Ham_diag}
 H=(\omega - \Omega )b_0^\dagger b_0 + \omega b_1^\dagger b_1 + (\omega + \Omega )b_2^\dagger b_2.
\end{eqnarray}
 Note that Eq.\ (\ref{newmodes}) can be inverted to give
\begin{eqnarray}
\label{newmodes_inv}
\nonumber &&a= \frac{1}{\sqrt{2}}\big(b_0 + b_2\big),\\
\nonumber &&\xi_1=\sqrt{\frac{k}{2N}}\big(b_2-b_0\big) + \sqrt{\frac{N-k}{N}} b_1,\\
 &&\xi_2=\sqrt{\frac{N-k}{2N}}\big(b_2-b_0\big) - \sqrt{\frac{k}{N}} b_1.
\end{eqnarray}

From the Hamiltonian, we easily find the time evolution of the operators $b_0,\ b_1$ and $b_2$ in
the Heisenberg picture, i.e.
\begin{eqnarray}
 && b_0(t)  =  e^{-i(\omega - \Omega)t}b_0(0),
\nonumber \\
 && b_1(t)  =  e^{-i\omega t}b_1(0),
\nonumber\\
 &&b_2(t) =  e^{-i(\omega + \Omega)t}b_2(0).
\end{eqnarray}
Combining these relations with (\ref{newmodes}) and (\ref{newmodes_inv}), we obtain the time evolution of the field and atomic operators:
\begin{eqnarray}
 && a(t) = e^{-i\omega t} \Big[ \cos (\Omega t) a(0) - i\sqrt{\frac{k}{N}}\sin (\Omega t) \xi_1(0)
 - i\sqrt{\frac{N-k}{N}}\sin (\Omega t) \xi_2 (0)\Big],
\nonumber \\
 &&\xi_1(t) =  e^{-i\omega t} \Big[ -i\sqrt{\frac{k}{N}}\sin (\Omega t) a(0) + \big(\frac{N-k}{N}
 + \frac{k}{N} \cos (\Omega t)\big) \xi_1(0)
\nonumber\\
 && \hspace{2cm}+ \frac{\sqrt{k(N-k)}}{N}\big(\cos (\Omega t) - 1 \big) \xi_2(0) \Big],
\nonumber \\
 &&\xi_2(t)= e^{-i\omega t} \Big[ -i\sqrt{\frac{N-k}{N}}\sin (\Omega t) a(0)
 + \frac{\sqrt{k(N-k)}}{N}\big(\cos (\Omega t) - 1 \big) \xi_1(0)
\nonumber\\
 && \hspace{2cm} + \big(\frac{k}{N} + \frac{N-k}{N} \cos (\Omega t)\big) \xi_2(0)  \Big].
\end{eqnarray}

Let us first consider the entanglement condition in Eq.\ (\ref{ineq2}),
and choose $A=\xi_1$ and $B=\xi_2$. The resulting inequality will tell us if the two groups of atoms are
entangled.  If the initial state is of the form
\begin{equation}
\ket{\Psi}=\ket{\psi}\ket{0,0},
\end{equation}
then the quantities at time $t$ appearing in the entanglement condition can be easily calculated, since only the terms involving the field operator at $t=0$ can be non-zero. In particular, we obtain
\begin{eqnarray}
 &&\langle \xi_1(t)\xi_2(t)\rangle
  =  -e^{-2i\omega t}
 \frac{\sqrt{k(N-k)}}{N}\sin^2 (\Omega t)\bra{\psi} (a(0))^2 \ket{\psi},
\nonumber \\
 &&\langle \xi^\dagger_1(t)\xi_1(t)\rangle =
 \frac{k}{N}\sin^2 (\Omega t)\bra{\psi}a^\dagger (0) a(0) \ket{\psi},
\nonumber \\
 &&\langle \xi_{2}^{\dagger}(t)\xi_2(t)\rangle  =
 \frac{N-k}{N}\sin^2 (\Omega t)\bra{\psi}a^\dagger (0) a(0) \ket{\psi},
\end{eqnarray}
hence for $t>0$, the entanglement condition becomes
\begin{equation}
 |\bra{\psi} (a(0))^2 \ket{\psi}|^2 > \bra{\psi}a^\dagger (0) a(0) \ket{\psi}^2 .
\end{equation}

We first note that a coherent state will not satisfy this condition.  That is not surprising, because
under the action of our Hamiltonian, an initial state with the field in a coherent state and the
atoms in their ground states will evolve into a product of coherent states in all three modes.
Such a state is clearly not entangled.  If the field is initially in a squeezed vacuum state
$\ket{\psi}=S(z)\ket{0}$
 the situation is very different. The entanglement condition reduces to
\begin{equation}
 \cosh^2 r >\sinh^2 r
\end{equation}
which is satisfied for any $r\neq 0$.  Therefore, squeezing in the initial field will lead to entanglement
between the two groups of atoms.

Let us now consider the entanglement condition in Eq (\ref{ineq1})
and again choose $A=\xi_1$ and $B=\xi_2$. If the initial state at $t=0$ is of the same form as before,
$\ket{\Psi}=\ket{\psi}\ket{0,0}$ ,
then the condition can be again greatly simplified as most of the terms on both sides of the equation are zero. We find that
\begin{eqnarray}
 \langle \xi_1^\dagger (t) \xi_2 (t)\rangle
 &=& \frac{\sqrt{k(N-k)}}{N}\sin^2(\Omega t)\bra{\psi}a^\dagger (0) a(0)\ket{\psi},
\nonumber  \\
 \langle \xi_1^\dagger (t) \xi_1 (t)\xi_2^\dagger (t) \xi_2 (t)\rangle
 &=& \frac{k(N-k)}{N^2}\sin^4(\Omega t)\Big[\bra{\psi}(a^\dagger (0) a(0))^2 \ket{\psi} \nonumber \\
 & &  - \bra{\psi}a^\dagger (0) a(0) \ket{\psi}\Big] ,
\end{eqnarray}
and the entanglement condition becomes
\begin{equation}
 \bra{\psi}a^\dagger (0) a(0)\ket{\psi}^2 >
 \bra{\psi}(a^\dagger (0) a(0))^2 \ket{\psi} - \bra{\psi}a^\dagger (0) a(0) \ket{\psi}.
\end{equation}
Expressed in terms of the photon number operator, $n=a^\dagger a $, this is just
 $\Delta^2 n(0) < \langle n(0) \rangle$, where $\Delta^2 n= \langle n^2 \rangle - \langle n \rangle^2$ is
the variance of $n$. Therefore, we see that there will be entanglement between the two groups of
atoms provided that the field is initially in a state with sub-Poissonian photon statistics.
Connections between nonclassical states and entanglement have been noted before.  For example, for
the output state of a beam splitter to be entangled, the input state must be nonclassical
\cite{kim}. The amount of  two-mode entanglement that can be produced by a single-mode field using
linear optics, auxiliary classical fields and ideal photodetectors has even been proposed as a
measure of the nonclassicality of a state \cite{asboth}.  Here we see yet another connection
between nonclassical states an entanglement in the dynamics of the Dicke model.

Light has, in fact, been used to entangle atomic ensembles, though this is often done by letting the
light interact with both sets of atoms and then measuring it \cite{duan}-\cite{julsgaard}.  The system we have considered here is a cartoon version of the one in the experiment \cite{julsgaard}, but it does
provide useful information in that it shows that certain kinds of nonclassical light can lead to atomic
entanglement.

\subsection{Two beam splitters}
Let us now consider a system of field modes that is closely related, at least in its description, to the
atom-field system we have just studied.  The system is depicted in Fig. 2.  It consists of three field modes
and two beam splitters.  One mode, mode $a$, passes through both beam splitters, while the other two,
modes $b$ and $c$, pass through just one. Our goal is to entangle modes $b$ and $c$ by making them interact with mode $a$ through the array of two beam splitters. Because this system closely resembles our previous example where a light mode was used to entangle two groups of atoms, it is  not surprising that the entanglement condition Eq.\ (\ref{ineq1}) again provides a usefull tool to derive
simple conditions on the input state of the ancillary mode $a$ that will result in modes $b$ and $c$
being entangled at the output.  This condition was applied to determine when the output of a single
beam splitter is entangled in \cite{hillery2}.

The system we are examining here was first studied in \cite{nagaj}.  In that paper a series of beam
splitters was used to model a reservoir, and, conditions for the entanglement of reservoir modes, which would correspond to our modes $b$ and $c$, were found.  However, in that paper only Gaussian states
were considered.

\begin{figure}
\centering
\includegraphics{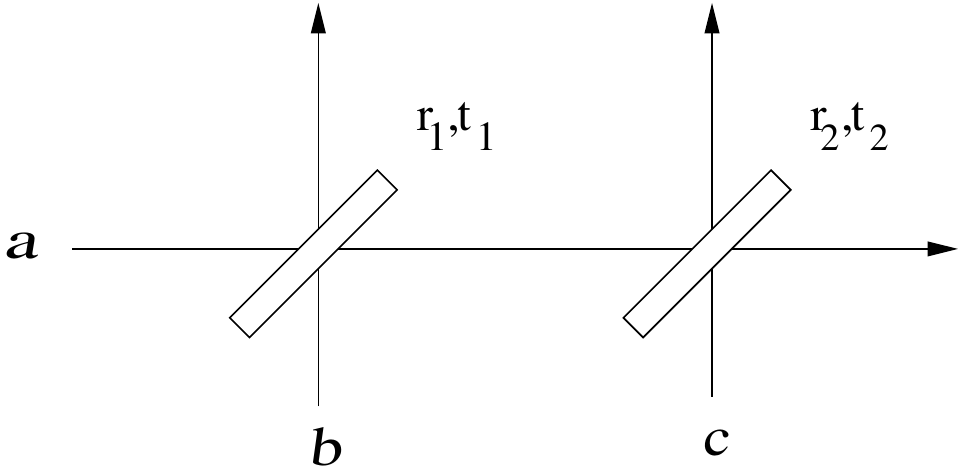}
\caption{Entangling modes $b$ and $c$ using the auxiliary mode $a$.} \label{fig:bs}
\end{figure}

We recall that a beam splitter acts on two input modes $a$ and $b$ as
\begin{eqnarray}
 a_{out} &=& t a_{in} + r b_{in}, \nonumber  \\
b_{out} &=& -r a_{in} + t b_{in},
\end{eqnarray}
where $r$ and $t$ are positive, and $r^2 + t^2=1$.
For beam splitters with transmittance and reflectance $(t_1,r_1)$ and $(t_2,r_2)$ respectively, the relationship between the output modes and the input modes is given by
\begin{eqnarray}
\label{inout}
\nonumber a_{out} &=& t_2(t_1 a_{in} + r_1 b_{in}) + r_2 c_{in}, \\
\nonumber b_{out} &=&  -r_1 a_{in} + t_1 b_{in}, \\
c_{out} &=& -r_2(t_1 a_{in} + r_1 b_{in}) + t_2 c_{in}.
\end{eqnarray}
Let us suppose for simplicity that modes $b$ and $c$ are initially in the vacuum, i.e.
$\ket{\Psi_{in}}=\ket{\psi}\ket{0}\ket{0}$. The entanglement condition in Eq.\ (\ref{ineq1}) for
$A=b_{out}$ and $B=c_{out}$ is
\begin{equation}
\label{cond_bs}
 |\langle b_{out}^\dagger c_{out} \rangle|^2 > \langle b_{out}^\dagger b_{out} c_{out}^\dagger c_{out} \rangle \,.
\end{equation}
Replacing  $b_{out}$ and $c_{out}$ by their expression Eq.\ (\ref{inout}) we find
\begin{eqnarray}
\nonumber \langle b_{out}^\dagger c_{out} \rangle &=&  t_1 r_1 r_2 \bra{\psi} a_{in}^\dagger a_{in} \ket{\psi},\\
\langle b_{out}^\dagger b_{out} c_{out}^\dagger c_{out} \rangle &=& (r_1 t_1 r_2)^2 \left[\bra{\psi} (a_{in}^\dagger a_{in})^2 \ket{\psi}  - \bra{\psi} a_{in}^\dagger a_{in} \ket{\psi}\right],
\end{eqnarray}
and the entanglement condition becomes, assuming that $r_1 t_1 r_2 \neq 0$,
\begin{equation}
\label{ent_cond_bs}
 \left[\langle (a_{in}^\dagger a_{in})^2 \rangle- \langle a_{in}^\dagger a_{in} \rangle^2\right]
< \langle a_{in}^\dagger a_{in} \rangle .
\end{equation}
i.e., $b$ and $c$ will be entangled provided that mode $a$ is initially in a state with sub-Poissonian statistics.

Now let us see what happens when we apply a more powerful entanglement condition.  Motivated by
considerations similar to those in Section V, we set $A=z_{1}b_{out}^{\dagger} + z_{2}b_{out}$ and
$B=c_{out}$ in Eq.\ (\ref{ineq1}).  We then find that a state is entangled if the $2\times 2$
matrix $M$,
where
\begin{eqnarray}
M_{11} & = & |\langle b_{out}c_{out}\rangle |^{2} - \langle b_{out}b^{\dagger}_{out}c^{\dagger}_{out}
c_{out}\rangle,  \nonumber \\
M_{12} & = & M_{21}^{\ast} = \langle b_{out}c_{out}\rangle \langle b_{out}c_{out}^{\dagger}\rangle
-\langle b_{out}^{2}c_{out}^{\dagger}c_{out}\rangle, \nonumber \\
M_{22} & = & |\langle b_{out}^{\dagger}c_{out}\rangle |^{2} - \langle b_{out}^{\dagger}b_{out}
c^{\dagger}_{out}c_{out}\rangle ,
\end{eqnarray}
has a positive eigenvalue, which will be the case if $|M_{12}|^{2}> M_{11}M_{22}$.  When we
express this condition in terms of the input operators, and assume that the $b$ and $c$ modes
are initially in the vacuum state, we find that the two-mode output state is entangled if
\begin{eqnarray}
 \label{ent_bs2} |\langle a^{2}_{in}\rangle \langle n\rangle - \langle na_{in}^{2}\rangle |^{2}  >
 [\langle n\rangle - (\Delta^{2} n) ]
 \left[ | \langle a_{in}^{2}\rangle |^{2} - \langle n^{2}\rangle +
 \left(1-\frac{1}{r_{1}^{2}}\right) \langle n\rangle \right] ,
\end{eqnarray}
where $n=a^{\dagger}_{in}a_{in}$.  Note that the Schwarz inequality implies that the second factor
on the right-hand side is negative, so that if the input field has sub-Poissonian statistics, then this
inequality is satisfied.  Therefore, this condition includes the one in Eq.\ (\ref{ent_cond_bs}).  There
are, however, states that satisfy the new condition but do not satisfy the old one.  For example, for
the state
\begin{equation}
|\psi\rangle = \left[ 1 - \left(\frac{1}{\sqrt{2}} + \epsilon \right)^{2} \right]^{1/2} |0\rangle
+  \left(\frac{1}{\sqrt{2}} + \epsilon \right) |2\rangle ,
\end{equation}
we find that $(\Delta^{2}n) -\langle n\rangle =- 2\sqrt{2} \epsilon$, to lowest order in
$\epsilon$.  On the other hand, Eq.\ (\ref{ent_bs2}) becomes, again keeping only lowest order
terms in $\epsilon$
\begin{equation}
\frac{1}{2}  >  -\sqrt{2} \epsilon .
\end{equation}
So, if $\epsilon < 0$ and small, then the state does not have sub-Poissonian statistics, but
Eq.\ (\ref{ent_bs2}) is satisfied, so the output will be entangled.  Therefore, Eq.\ (\ref{ent_bs2}) is
a stronger condition than Eq.\ (\ref{ent_cond_bs}).

\section{Local uncertainty relations and two modes}
So far we have been concentrating on the consequences of the inequalities in Eqs.\ (\ref{ineq1})
and (\ref{ineq2}).  Needless to say, it is possible to derive other entanglement conditions involving
non-hermitian operators.  One way, which we will briefly explore here is to apply a variant of the local
uncertainty entanglement condition due to Hofmann and Takeuchi \cite{hofmann}.
Let us start by considering two modes.  Now suppose that $A$ is an operator on
mode $a$ and $B$ is an operator on mode $b$.  We want to first find an expression for
$\langle(A^{\dagger} + B^{\dagger}) (A+B)\rangle - |\langle A+B\rangle |^{2}$ for a separable state
\begin{equation}
\rho = \sum_{k} p_{k}\rho^{(a)}_{k}\otimes \rho^{(b)}_{k} .
\end{equation}
We first note that for an operator $D$ and a density matrix
\begin{equation}
\rho = \sum_{k} p_{k}\rho_{k},
\end{equation}
we have that
\begin{eqnarray}
\langle D^{\dagger}D\rangle - |\langle D\rangle |^{2}& = &
\sum_{k} p_{k}[\langle D^{\dagger}D\rangle_{k}
- \langle D^{\dagger}\rangle_{k} \langle D\rangle_{k} + (\langle D^{\dagger}\rangle_{k}
-\langle D^{\dagger}\rangle )(\langle D\rangle_{k} - \langle D\rangle )] \nonumber \\
 & \geq & \sum_{k} p_{k} ( \langle D^{\dagger}D\rangle_{k}
- \langle D^{\dagger}\rangle_{k} \langle D\rangle_{k} ) ,
\end{eqnarray}
where expectation values with the subscript $k$ denote the expectation value with respect to
$\rho_{k}$.  If we now apply this to $D=A+B$ with the separable density matrix above, we find
that
\begin{equation}
\label{onepermode}
\langle(A^{\dagger} + B^{\dagger}) (A+B)\rangle - |\langle A+B\rangle |^{2} \geq \sum_{k} p_{k}
(\langle A^{\dagger}A\rangle_{k} - | \langle A\rangle_{k}|^{2} + \langle B^{\dagger}B\rangle_{k}
- |\langle B\rangle_{k}|^{2} ) .
\end{equation}
This condition can easily be extended to the case in which we have more than one operator for
each subsystem.
If we have M operators for mode $a$, $A_{j}$ and $M$ for mode $b$, $B_{j}$, then the above
inequality generalizes to
\begin{eqnarray}
\label{twopermode}
\sum_{j=1}^{M}\langle(A_{j}^{\dagger} + B_{j}^{\dagger}) (A_{j}+B_{j})\rangle
- |\langle A_{j}+B_{j}\rangle |^{2} \geq \nonumber \\
\sum_{k} p_{k} \sum_{j=1}^{M}
(\langle A_{j}^{\dagger}A_{j}\rangle_{k} - | \langle A_{j}\rangle_{k}|^{2}
+ \langle B_{j}^{\dagger}B_{j}\rangle_{k} - |\langle B_{j}\rangle_{k}|^{2} ) .
\end{eqnarray}
These two inequalities will hold for all separable density matrices, so if a state violates them, it
must be entangled.

As a simple example of an entanglement condition that can be derived from Eq.\ (\ref{onepermode}),
let us set $A=a$ and $B=b^{\dagger}$.  We then have that
\begin{eqnarray}
\langle (a^{\dagger} + b)(a+b^{\dagger})\rangle - |\langle a+b^{\dagger}\rangle |^{2}  & \geq &
\sum_{k} p_{k}(\langle a^{\dagger}a\rangle_{k} -|\langle a \rangle_{k}|^{2}
+ \langle bb^{\dagger}\rangle_{k} - |\langle b \rangle_{k}|^{2} )  \nonumber  \\
 & \geq & 1 .
 \end{eqnarray}
 If this inequality is violated by a state, that state is entangled.  A two-mode squeezed vacuum
 state,
 \begin{equation}
 |\psi_{sq}\rangle = e^{r(a^{\dagger}b^{\dagger}-ab)}|0\rangle ,
 \end{equation}
will violate this inequality.  We find that for this state
\begin{equation}
\langle (a^{\dagger} + b)(a+b^{\dagger})\rangle - |\langle a+b^{\dagger}\rangle |^{2} = e^{-2r} ,
\end{equation}
so that for $r>0$, the condition is violated, and we can conclude that the state is entangled.  It
should be noted that this condition can also be derived from the analysis given by Shchukin
and Vogel \cite{shchukin}.

A very similar condition can be derived for atom-field entanglement.  If we have $N$ two-level atoms
represented by collective spin operators $S^{+}$, $S^{-}$, and $S^{z}$, then setting $A=a^{\dagger}$
and $B=J_{+}$ we find that for a separable state
\begin{equation}
\langle (a+S^{-})(a^{\dagger}+ S^{+})\rangle -|\langle a + S^{-}\rangle |^{2} \geq 1 ,
\end{equation}
so that any state that violates this inequality must be entangled.  A very simple example is the state,
for $N=1$
\begin{equation}
|\psi\rangle = \cos\theta |e\rangle |0\rangle + \sin\theta |g\rangle |1\rangle ,
\end{equation}
for $0 < \theta < \pi /4$.

\section{Conclusion}
We have presented a number of entanglement conditions that employ non-hermitian operators.  While
these conditions follow from the partial transpose condition, they are much simpler to use.  It is only
necessary to compute a small number of correlation functions rather than to have to diagonalize the
partially transposed density matrix for the entire system.  We used these conditions to show the
presence of entanglement in a number of systems of interest in quantum optics, including
Jaynes-Cummings model and the Dicke model.
The conditions are sufficiently flexible that we can study
entanglement between modes, entanglement between a field mode and an atom, and entanglement
between groups of atoms.  Most of these conditions are invariant under some set of local unitary
transformations.  For field modes we considered local Gaussian operations.  By making the conditions
invariant under a set of local operations, we are able to make them able to detect entanglement in
a wider class of states.

\section*{Acknowledgments}
Julien Niset acknowledges support from the Belgian American Educational Foundation,
and Ho Trung Dung acknowledges support from the Vietnam Education Foundation.  This
research was partially supported by the National Science Foundation under grant PHY-0903660.

\section*{Appendix}
Here we present the details needed for the calculations involving the Tavis-Cummings model.  As
we noted, the Hilbert space splits up into a direct sum of invariant subspaces, $\Hi^{(n)}$ ,which are characterized by the total excitation number, $n$.
$\Hi^{(0)}$ is spanned by $\{\ket{0,g,g}\}$ so the ground state of the system is
\begin{equation}
 \ket{\psi_0^{(0)}}=\ket{0,g,g}
\end{equation}
with eigenvalue
\begin{equation}
 E_0^{(0)}=\bra{0,g,g}\Hi \ket{0,g,g}=-\omega.
\end{equation}

$\Hi^{(1)}$ is spanned by $\{ \ket{1,g,g},\ket{0,e,g},\ket{0,g,e}\}$, and the Hamiltonian can be written in this basis as
\begin{equation}
H^{(1)}=\left(
\begin{array}{ccc}
0 & \kappa & \kappa \\
\kappa & 0 & 0 \\
\kappa & 0 & 0
\end{array}
\right).
\end{equation}
This Hamiltonian has three eigenvalues:
\begin{eqnarray}
\nonumber E_0^{(1)}&=&-\kappa \sqrt{2},\\
\nonumber E_1^{(1)}&=&0,\\
E_2^{(1)}&=&\kappa \sqrt{2},
\end{eqnarray}
to which we associate the three eigenvectors
\begin{eqnarray}
\label{nequal1}
\nonumber\ket{\psi_0^{(1)}}&=& \frac{1}{\sqrt{2}}\ket{1,g,g}
 - \frac{1}{\sqrt{2}}\big[\frac{\ket{0,e,g} + \ket{0,g,e}}{\sqrt{2}}\big],\\
\nonumber\ket{\psi_1^{(1)}}&=& \frac{1}{\sqrt{2}}\big[ \ket{0,e,g} - \ket{0,g,e} \big],\\
\ket{\psi_2^{(1)}}&=& \frac{1}{\sqrt{2}}\ket{1,g,g}
 + \frac{1}{\sqrt{2}}\big[\frac{\ket{0,e,g} + \ket{0,g,e}}{\sqrt{2}}\big].
\end{eqnarray}

$\Hi^{(n)}$ is spanned by $\{ \ket{n,g,g},\ket{n-1,e,g},\ket{n-1,g,e},\ket{n-2,e,e}\}$, and the Hamiltonian can be written in this basis as
\begin{equation}
H^{(n)}=\left(
\begin{array}{cccc}
\omega (n-1) & \kappa \sqrt{n} & \kappa \sqrt{n} & 0 \\
\kappa \sqrt{n} & \omega (n-1) & 0 & \kappa \sqrt{n-1} \\
\kappa \sqrt{n} & 0 & \omega (n-1) & \kappa \sqrt{n-1} \\
0 & \kappa \sqrt{n-1} & \kappa \sqrt{n-1} & \omega (n-1)
\end{array}
\right).
\end{equation}
This Hamiltonian has four eigenvalues:
\begin{eqnarray}
\nonumber E_0^{(n)}&=& \omega (n-1) - \kappa \sqrt{2(2n-1)},\\
\nonumber E_{1,2}^{(n)}&=& \omega (n-1),\\
 E_3^{(n)}&=& \omega (n-1) + \kappa \sqrt{2(2n-1)},
\end{eqnarray}
to which we can associate the four eigenvectors
\begin{eqnarray}
\nonumber\ket{\psi_0^{(n)}}&=& \sqrt{\frac{n}{2(2n-1)}}\ket{n,g,g} - \frac{1}{\sqrt{2}}\big[ \frac{\ket{n-1,e,g}
 + \ket{n-1,g,e}}{\sqrt{2}}\big],  \\
& & + \sqrt{\frac{n-1}{2(2n-1)}}\ket{n-2,e,e}, \nonumber  \\
\nonumber\ket{\psi_1^{(n)}}&=& \frac{1}{\sqrt{2}}\big[ \ket{n-1,e,g} - \ket{n-1,g,e} \big],\\
\nonumber\ket{\psi_2^{(n)}}&=& \sqrt{\frac{n-1}{2n-1}}\ket{n,g,g}-\sqrt{\frac{n}{2n-1}}\ket{n-2,e,e},\\
\ket{\psi_3^{(n)}}&=& \sqrt{\frac{n}{2(2n-1)}}\ket{n,g,g}
 + \frac{1}{\sqrt{2}}\big[ \frac{\ket{n-1,e,g} + \ket{n-1,g,e}}{\sqrt{2}}\big] \nonumber \\
 & & + \sqrt{\frac{n-1}{2(2n-1)}}\ket{n-2,e,e}.
\end{eqnarray}

Suppose that at time $t=0$, the field contains $n$ excitations while both atoms are in their ground
state, i.e.
\begin{eqnarray}
 \ket{\Psi(0)}&=& \ket{n,g,g}\\
\nonumber &=& \sqrt{\frac{n}{2(2n-1)}}\ket{\psi_0^{(n)}} +\sqrt{\frac{n-1}{2n-1}}\ket{\psi_2^{(n)}} 
 +\sqrt{\frac{n}{2(2n-1)}}\ket{\psi_3^{(n)}}.
\end{eqnarray}
Note that this expression is valid for any $n>0$, provided that the appropriate eigenvectors are chosen.
 In particular, when $n=1$, $\ket{\psi_3^{(n)}}$ corresponds to $\ket{\psi_2^{(1)}}$ as defined
  in Eq. (\ref{nequal1}).

After a time $t$, the state of the system will have evolved according to
\begin{eqnarray}
\label{psit_2}
\nonumber \ket{\Psi(t)}&=& e^{-iH^{(n)}t}\ket{\Psi(0)}\\
\nonumber&=& \sqrt{\frac{n}{2(2n-1)}}e^{-iE_0^{(n)}t}\ket{\psi_0^{(n)}}+\sqrt{\frac{n-1}{2n-1}}e^{-iE_2^{(n)}t}\ket{\psi_2^{(n)}} \\
& & +\sqrt{\frac{n}{2(2n-1)}}e^{-iE_3^{(n)}t}\ket{\psi_3^{(n)}} \nonumber \\
\nonumber&=& e^{-i\omega (n-1)t}\Big[ \frac{1}{2n-1}\big( n \cos(\Omega t)+n-1\big)\ket{n,g,g} \\
& & + \frac{\sqrt{n(n-1)}}{(2n-1)}\big( \cos(\Omega t)-1 \big)\ket{n-2,e,e} \nonumber \\
 & & - i \sqrt{\frac{n}{2n-1}}\sin(\Omega t)\big[  \frac{\ket{n-1,e,g} +
 \ket{n-1,g,e}}{\sqrt{2}}\big]\Big],
\end{eqnarray}
where $\Omega=\kappa \sqrt{2(2n-1)}$.


\begin{thebibliography}{99}
\bibitem{horodecki}R.\ Horodecki, P.\ Horodecki, M.\ Horodecki, and K.\ Horodecki, Rev.\ Mod.\ Phys.
{\bf 81}, 865 (2009).
\bibitem{guehne} O.\ G\"{u}hne and G.\ Toth, Physics Reports {\bf 474}, 1 (2009).
\bibitem{simon} R.\ Simon, Phys.\ Rev.\ Lett. {\bf 84},  2726 (2000).
\bibitem{duan} L.\ -M.\ Duan, G.\ Giedke, J.\ I.\ Cirac, and P.\ Zoller, Phys.\ Rev.\ Lett. {\bf 84},
2722 (2000).
\bibitem{mancini} S.\ Mancini, V.\ Giovannetti, D.\ Vitali, and P.\ Tombesi, Phys.\ Rev.\ Lett. {\bf 88},
120401 (2002).
\bibitem{agarwal} G.\ S.\ Agarwal and A.\ Biswas, New Journal of Physics {\bf 7}, 211 (2005).
\bibitem{hillery} Mark Hillery and M. Suhail Zubairy, Phys.\ Rev.\ Lett.\ {\bf 96}, 050503 (2006).
\bibitem{shchukin} E.\ Shchukin and W.\ Vogel, Phys.\ Rev.\ Lett.\ {\bf 95}, 230502 (2005).
\bibitem{toth} Geza Toth, Christoph Simon, and Juan Ignacio Cirac, Phys.\ Rev.\ A {\bf 68},
062310 (2003).
\bibitem{nha} Hyunchul Nha and Jaewan Kim, Phys.\ Rev.\ A {\bf 74}, 012317 (2006).
\bibitem{buzek} V. Bu\v{z}ek, M. Orszag, and M. Rosko, Phys. Rev. Lett. \textbf{94}, 163601 (2005).
\bibitem{kim}M.\ S.\ Kim, W.\ Son, V.\ Bu\v{z}ek, and P.\ L.\ Knight, Phys.\ Rev.\ A {\bf 65} 032323
(2002).
\bibitem{asboth} J.\ K.\ Asboth, J.\ Casamiglia, and H.\ Ritsch, Phys.\ Rev.\ Lett. {\bf 94}, 173602
(2005).
\bibitem{duan} L.\ -M.\ Duan, J.\ I.\ Cirac, P.\ Zoller, and E.\ S.\ Polzik, Phys.\ Rev.\ Lett. {\bf 85}, 5643
(2000).
\bibitem{kuzmich} A.\ Kuzmich and E.\ Polzik, Phys.\ Rev.\ Lett. {\bf 85}, 5639 (2000).
\bibitem{hammerer} For a review see K.\ Hammerer, A.\ S.\ Sorensen, and E.\ S.\ Polzik,
quant-ph/0807.3358.
\bibitem{julsgaard} B.\ Julsgaard, A. Kozhekin, and E. Polzik, Nature {\bf 413}, 6854 (2001).
\bibitem{hillery2} Mark Hillery and M. Suhail Zubairy, Phys.\ Rev.\ A {\bf 74}, 032333 (2006).
\bibitem{nagaj} Daniel Nagaj, Peter \v{S}telmachovi\v{c}, Vladimir Bu\v{z}ek, and Myungshik
Kim, Phys.\ Rev.\ A {\bf 66}, 062307 (2002).
\bibitem{hofmann} Holger Hofmann and Shigeki Takeuchi, Phys.\ Rev.\ A {\bf 68}, 032103 (2003).
\end{thebibliography}
\end{document}